\documentclass[draftcls,onecolumn,journal]{IEEEtran}

\ifCLASSINFOpdf
\else
   \usepackage[dvips]{graphicx}
\fi
\usepackage{url}

\usepackage{graphicx}
\usepackage{amsmath, amsfonts, amssymb}
\usepackage[colorlinks=true, allcolors=blue]{hyperref}

\usepackage[ruled, lined, linesnumbered, longend]{algorithm2e}

\usepackage{color}
\usepackage{xcolor}

\definecolor{webgreen}{rgb}{0,.5,0} 
\definecolor{lightgreen}{rgb}{0.6,.8,0.6} 
\definecolor{lightgray}{rgb}{0.78,0.78,0.78} 

\SetKwInOut{KwIn}{Input}
\SetKwInOut{KwOut}{Output}

\SetCommentSty{mycommfont}
\newcommand{\mycomment}[1]{\tcp{#1}}

\usepackage{ulem}

\usepackage{soul}

\renewcommand{\emph}{\textit}

\usepackage{multirow}
\usepackage{array}
\usepackage{booktabs}
\usepackage{xfrac}
\usepackage{color}

\usepackage{comment}
\usepackage{hyperref}

\usepackage[compress]{cite}

\usepackage{graphicx}
\usepackage{tikz}
\usetikzlibrary{shapes.misc}
\tikzset{cross/.style={cross out, draw, 
         minimum size=2*(#1-\pgflinewidth), 
         inner sep=0pt, outer sep=0pt}}
 \usetikzlibrary{calc}         
\usepackage{pgfplots}
\usetikzlibrary{3d}

\def\argmin{\mathop{\mathsf{arg\,min}}} 

\def\lim{\mathop{\mathsf{lim}}} 
\def\min{\mathop{\mathsf{min}}}

\def\log{\mathrm{log}}

\newcommand\norm[1]{\left\lVert#1\right\rVert}

\newcommand{\bsC}{\boldsymbol{C}}

\newcommand{\bsF}{\boldsymbol{F}}

\newcommand{\bsy}{\boldsymbol{y}}
\newcommand{\bsY}{\boldsymbol{Y}}

\newcommand{\bsv}{\boldsymbol{v}}
\newcommand{\bsV}{\boldsymbol{V}}

\newcommand{\bss}{\boldsymbol{s}}

\newcommand{\bsS}{\boldsymbol{S}}

\newcommand{\bsQ}{\boldsymbol{Q}}

\def\R{\mathbb{R}}

\begin{document}

\title{Online simplex-structured matrix factorization }


\author{Hugues Kouakou, José Henrique de Morais Goulart, Raffaele Vitale, Thomas Oberlin, \\
David Rousseau, Cyril Ruckebusch, and Nicolas Dobigeon,~\IEEEmembership{Senior Member,~IEEE}

\thanks{This work was supported by the ANR IMAGIN Research Project under Grant No. ANR-21-CE29-0007. }
\thanks{H. Kouakou, J. H. de Morais Goulart, and N. Dobigeon are with Univ. de Toulouse, IRIT/INP-ENSEEIHT, 31071 Toulouse, France (email: \{hugues.kouakou, henrique.goulart, nicolas.dobigeon\}@irit.fr).}
\thanks{R. Vitale and C. Ruckebusch are with Univ. Lille, CNRS, LASIRE, 59000 Lille, France (email: \{raffaele.vitale, Cyril.ruckebusch\}@univ-lille.fr).}
\thanks{T. Oberlin is with Univ. de Toulouse, ISAE-SUPAERO, 31400 Toulouse, France (email: thomas.oberlin@isae-supaero.fr).}
\thanks{D. Rousseau is with Univ. d’Angers, LARIS, UMR IRHS INRAe, 49000 Angers, France (email: david.rousseau@univ-angers.fr).}}

\maketitle

\begin{abstract}
Simplex-structured matrix factorization (SSMF) is a common task encountered in signal processing and machine learning. Minimum-volume constrained unmixing (MVCU) algorithms are among the most widely used methods to perform this task. While MVCU algorithms generally perform well in an offline setting, their direct application to online scenarios suffers from scalability limitations due to memory and computational demands. To overcome these  limitations, this paper proposes an approach which can build upon any off-the-shelf MVCU algorithm to operate sequentially, i.e., to handle one observation at a time. The key idea of the proposed method consists in updating the solution of MVCU only when  necessary, guided by an online check of the corresponding optimization problem constraints. It only stores and processes observations identified as informative with respect to the geometrical constraints underlying SSMF. We demonstrate the effectiveness of the approach when analyzing synthetic and real datasets, showing that it achieves estimation accuracy comparable to the offline MVCU method upon which it relies, while significantly reducing the computational cost.
\end{abstract}

\begin{IEEEkeywords}
Simplex-structured matrix factorization, online processing, low-rank approximation.
\end{IEEEkeywords}

\IEEEpeerreviewmaketitle

\section{Introduction}\label{sec:introduction}

Non-negative matrix factorization is a widely used method in signal processing and machine learning for extracting latent low-dimensional representations from high-dimensional, non-negative data \cite{fu2019nonnegative, gillis2020nonnegative, gligorijevic2018non, huang2013non, lee2009semi, liu2025simplex}. This is commonly achieved by approximating the data matrix \(\bsY \in \R_{+}^{L\times T}\) as
\begin{equation} \label{model}
\bsY \approx \bsS\bsC,
\end{equation}
where \( \bsS \in \R_{+}^{L\times K} \) contains $K$ so-called basis vectors and \(\bsC \in \R_{+}^{K\times T}\) is the matrix of the corresponding coefficients, with \( K \ll L\). This formulation arises in several fields such as chemometrics \cite{de2021multivariate, mansouri2024current}, where \(\bsY\) contains spectra measured at $L$ wavelengths and \(\bsS\) encodes the spectral signatures of the $K$ main components, and topic modeling \cite{abdelrazek2023topic}, where \(\bsY\) encodes document-term frequencies over a vocabulary of size $L$, and \(\bsS\) specifies the average frequency of each word within each of the $K$ latent topics.  When the columns of \(\bsC\) are constrained to lie in the standard simplex, i.e., being nonnegative and summing to one, this formulation leads to what is commonly known as simplex-structured matrix factorization (SSMF). It admits a simple geometric interpretation, in which the columns of \(\bsY\) lie within a simplex whose vertices correspond to the columns of \(\bsS\). Numerous works have been dedicated to the development of SSMF algorithms  \cite{chan2009convex, bioucas2009variable, qu2015subspace, lin2018maximum, wu2021probabilistic, granot2023probabilistic} or to the characterization of the SSMF problem under additional assumptions \cite{gillis2014robust, fu2015robustness, abdolali2021simplex, thanh2023bounded}.

In parallel to these methodological and theoretical advances, recent research efforts have been made in the spectroscopy and microscopy communities in order to set up new acquisition schemes to mitigate shortcomings inherent to the conventional experimental protocols, such as long acquisition times and the risk of damaging biological samples. For instance, the authors in \cite{coic2023assessment,gilet2023superpixels}, proposed a smart protocol dedicated to Raman spectroscopy that selectively targets the ``most informative" spectral pixels \cite{ghaffari2019essential}. Similar developments, notably sparse sampling strategies, have also been explored in scanning electron microscopy \cite{dahmen2019adaptive,dahmen2016feature}. These developments call for a more computationally efficient processing of incoming measurements (or observations), which would open the door to a real-time analysis. The work reported in this paper precisely aims at meeting this objective, i.e., designing an SSMF method suitable to be combined with these innovative protocols.

Recent studies have focused on accelerating SSMF by leveraging either hardware resources for efficient implementation of the method itself \cite{wu2021recent}, or by performing dimensionality and data reductions before conducting SSMF  \cite{liu2012analysis}. However, these accelerations have been essentially envisioned to fasten SSMF in large-scale data settings, typically in an offline manner. Conversely, several methods have been specifically designed to perform online processing, where data is handled sequentially (i.e., one observation at a time) and the matrix \(\bsS\) is updated accordingly. One popular method is online dictionary learning (ODL), where the estimation task is formulated as a stochastic optimization problem, and the cost function is approximated using mini-batches or a single observation at a time by aggregating information from past observations \cite{mairal2010online}. The method proposed in \cite{nus2020admm} updates \(\bsS\) based on batches of observations, each corresponding to a line of contiguous hyperspectral pixels acquired by a pushbroom sensor. Although this approach processes data row by row, it is not suitable to the acquisition protocols discussed above. A Kalman filter (KF)-based approach has been recently proposed to dynamically estimate \(\bsS\) \cite{kouakou2024fly}. However, Kalman filtering is known to be sensitive to observation noise and rely on strong assumption regarding the model of observation and the model of the hidden state (the vertices to be estimated), which significantly limits its performance in practice.

In this work, we address these limitations by proposing a more robust and computationally efficient method for online SSMF. Specifically, we introduce an online counterpart of MVCU, by building upon any existing off-the-shelf MVCU-based SSMF algorithm. Our approach adopts an online strategy that selects relevant observations by exploiting the geometric constraints underlying the SSMF problem.

\section{Minimum-volume constrained unmixing: background and online formulation}

The SSMF problem assumes that the observations lie within a $K-$dimensional linear subspace, in consonance with \eqref{model}. In the sequel of this letter, we assume the dimension $K$ is known and, without loss of generality, that \(\bsY\) has been projected onto this subspace. Thus, with a slight abuse of notation, we denote its reduced version by \(\bsY \in \R^{K\times T}\). 
Among the methods performing SSMF, MVCU algorithms such as minimum volume enclosing simplex (MVES) \cite{chan2009convex}  and minimum volume simplex analysis (MVSA) \cite{li2008minimum}  aim at solving the following nonconvex optimization problem \vspace{-0.10cm}
\begin{equation*}
    \mathcal{P}_{T}:\; \min_{\bsS}  \text{vol}(\bsS)  \quad \text{s.t.} \quad \bsS^{-1}\bsY \geq \mathbf{0} \quad \text{and} \quad \mathbf{1}^\top_K\bsS^{-1}\bsY = \mathbf{1}^\top_T \vspace{-0.10cm}
\end{equation*}
where \(\mathbf{1}_N\) is an $N$-dimensional column vector with all entries equal to $1$. Most MVCU algorithms use the determinant as a measure of the volume of the simplex whose vertices are the columns of \(\bsS\) \cite{chan2009convex, li2008minimum, shi2019endmember, ambikapathi2010robust, arngren2011unmixing}, since these quantities are proportional. 
Motivated by the online setting underlying this work, let us consider a set $\bsY_{\leq  t}=[\bsy_1,\ldots,\bsy_t]$ of $t\leq T$ observations  to be analyzed under the prism of SSMF. In this context, the conventional SSMF problem $\mathcal{P}_{T}$  addressed by MVCU is reformulated as a sequence of $T$ intermediate SSMF problems  $\mathcal{P}_{\leq t}$ ($t\leq T$) such that $\mathcal{P}_{\leq T} =\mathcal{P}_{T}$ and defined by \vspace{-0.10cm}
\begin{equation*}
   \mathcal{P}_{\leq t}:\;  \min_{\bsQ_t} \;-\log\left( \lvert \det(\bsQ_t) \rvert \right) 
\;  \text{s.t.} \; \forall j\leq t, \; 
            \bsy_j \in \mathcal{S}(\bsQ_t) \vspace{-0.10cm}
\end{equation*}
where \vspace{-0.10cm}
\begin{equation*}
    \mathcal{S}(\bsQ) = \left\{\bsy \in \mathbb{R}^K;\ g(\bsQ \bsy)\ge 0 \; \text{and} \;  h(\bsQ \bsy) = 0\right\} \vspace{-0.10cm}
\end{equation*}
defines the simplex whose vertices are the columns of $\bsS = \bsQ^{-1}$. The nonnegativity and sum-to-one constraints underlying SSMF are encoded by the two functions $g(\bsv) = \min_i [\bsv]_i$ and $h(\bsv) = 1 - \mathbf{1}_K^\top \bsv$. 

In an online SSMF context, we seek to update the  vertices of the enclosing simplex (i.e., the columns of $\hat{\bsS}_{t-1} = {\hat\bsQ}_{t-1}^{-1}$)  after receiving each incoming new observation $\bsy_t$ and given a current estimate $\hat{\bsQ}_{t-1}$.  In the sequel, we assume that we can build on an existing (off-the-shelf) MVCU method designed to solve the SSMF problem $\mathcal{P}_{\leq t}$. Since most of the state-of-the-art MVCU methods are iterative algorithms, the output of this algorithm will be denoted as $\hat{\bsQ}_{t}=\textsf{MVCU}(\bsY_{\leq t},\hat{\bsQ}_{t-1})$ where $\hat{\bsQ}_{t-1}$ serves as the initialization of the considered MVCU algorithm. The next section details the strategy proposed to enable a computationally efficient implementation of an MVCU algorithm within the framework of online SSMF (oSSMF). 

\section{Proposed oSSMF method}
This section presents the proposed oSSMF method step-by-step. We first identify in Section~\ref{sec: naive} the observations that call for updating the current estimate of the simplex. Without any refinement, the resulting algorithmic procedure would be too computationally demanding to be implemented in an online setting. Thus  Section~\ref{sec: relevant_observations} introduces an iterative selection of observations identified as relevant for the SSMF problem. Section~\ref{sec: ReducingRedundancy} completes the proposed method by further reducing redundancy in the observations. Details regarding the practical implementation and a algorithmic sketch are reported in Section~\ref{sec: pseudocode}.

\subsection{Avoiding unnecessary updates}
\label{sec: naive}
Let $\Hat{\bsQ}_{t-1}$ denote the current solution of the SSMF problem $\mathcal{P}_{\leq t-1}$.  Once a new observation $\bsy_t$ has to be handled, one wants then to solve the problem $\mathcal{P}_{\leq t}$. It clearly appears that if $\bsy_t \in \mathcal{S}(\bsQ_{t-1})$, the problems \(\mathcal{P}_{\leq t-1}\) and \(\mathcal{P}_{\leq t}\) share the same solution. In other words, no update is required since we have $\textsf{MVCU}(\bsY_{\leq t},\hat{\bsQ}_{t-1}) = \textsf{MVCU}(\bsY_{\leq {t-1}},\hat{\bsQ}_{t-2})$. Otherwise, if the newly acquired observation \(\bsy_t\) violates at least one of the current constraints, i.e., $\bsy_t \not \in \mathcal{S}(\hat{\bsQ}_{t-1})$, then an update is triggered. The selected MVCU method can be initialized using the current estimate, and we run \(\textsf{MVCU}(\bsY_{\leq t},\hat{\bsQ}_{t-1})\). Even if it avoids running MVCU again when it is not necessary, this naive implementation results in an increasing computational cost as the number of observations $t$ grows in $\bsY_{\leq t}$. This computational burden can be mitigated by identifying a subset of relevant observations to be considered during SSMF.

\subsection{Identifying relevant observations}
\label{sec: relevant_observations}
Instead of running the MVCU algorithm on the whole set of observations $\bsY_{\leq t}$ acquired so far, we propose to iteratively identify a subset $\bsV_t \subseteq \bsY_{\leq t}$ of relevant observations that should be kept after receiving the observation $\bsy_t$. 
This iterative selection depends on whether the observation $\bsy_t$ has called for updating $\hat{\bsQ}_{t-1}$ (see Section \ref{sec: naive}). The two cases are discussed below.\\

\noindent \textbf{Case 1: $\bsy_t \not\in \mathcal{S}(\bsQ_{t-1})$} -- When the newly acquired observation has called for updating the current estimate $\hat{\bsQ}_{t-1}$, the algorithm \(\textsf{MVCU}(\bsV_{t-1} \cup \left\{\bsy_t\right\},\hat{\bsQ}_{t-1})\) is run on the set of current relevant observations $\bsV_{t-1}$ previously identified and augmented with the new observation $\bsy_t$. Given the new estimate $\hat{\bsQ}_{t}$ such that $\mathcal{S}(\hat{\bsQ}_{t-1}) \neq \mathcal{S}(\hat{\bsQ}_{t})$, a first set of relevant observations are those that do not satisfy the updated constraints, i.e., $\forall \bsy_j \in \bsV_{t-1},\  \bsy_j \not\in \mathcal{S}(\hat{\bsQ}_{t})$, since they will surely contribute to updating the new MVCU solution $\hat{\bsQ}_{t}$ after receiving the future observation $\bsy_{t+1}$. Moreover, among the observations belonging to the set $\bsV_{ t-1} \cup \left\{\bsy_t\right\}$, we are particularly interested in those which activate the constraints underlying the problem $\mathcal{P}_t$, i.e., those belonging to the simplex facets $\mathcal{F}(\hat{\bsQ}_t) \subset \mathcal{S}(\hat{\bsQ}_t)$ where
\begin{equation*}
  \mathcal{F}(\bsQ)  = \left\{\bsy \in \mathcal{S}(\bsQ);\ g(\bsQ \bsy) = 0 \right\}.
\end{equation*}
As a consequence, in this first case, after handling the observation $\bsy_t$ and computing $\hat{\bsQ}_t$, the set of relevant observations is updated as
\begin{equation}\label{eq:updating_rule1_strict}
    \bsV_{t} = \left\{\bsV_{t-1} \cup \left\{\bsy_t\right\}\right\} \cap \left\{\mathcal{S}(\hat{\bsQ}_t)^\mathrm{c} \cup \mathcal{F}(\hat{\bsQ}_t)\right\}
\end{equation}
where $\mathcal{A}^\mathrm{c}$ denotes the complement of the set $\mathcal{A}$.\\

\noindent \textbf{Case 2: $\bsy_t \in \mathcal{S}(\bsQ_{t-1})$} -- When the newly available observation $\bsy_t$ has not required running MVCU (i.e., $\hat{\bsQ}_{t} = \hat{\bsQ}_{t-1}$), it may be considered as a relevant observation only if it activates the constraints, i.e., it belongs to one of the facet of the currently estimated simplex. The resulting updating rule finally reads as
\begin{equation}\label{eq:updating_rule2_strict}
    \bsV_{t} = \bsV_{t-1} \cup \left\{\left\{\bsy_t\right\} \cap \mathcal{F}(\bsQ_t)\right\}.
\end{equation}

\subsection{Reducing redundancy}
\label{sec: ReducingRedundancy}
To further reduce the computational complexity of the proposed oSSMF method, we propose to constrain the updating rule \eqref{eq:updating_rule2_strict} of the relevant observations when  $\bsy_t \in \mathcal{F}({\bsQ_{t-1}})$. Indeed, storing \(\bsy_t\) by default as suggested by \eqref{eq:updating_rule2_strict} may lead to many redundant points that are not useful for solving the SSMF problem. An observation \(\bsy_t \in \mathcal{F}({\bsQ_{t-1}})\) is considered as non-redundant for the SSMF problem if it is sufficiently distant from all other relevant observations in \(\bsV_{t-1}\) located on the same facet. The facet of a simplex containing $\bsy_t$ and whose vertices are the columns of $\bsS=\bsQ^{-1}$ can be easily identified as the set
\begin{equation*}
    {\mathcal{A}}({{\bsQ}},\bsy_t) = \left\{\bsy \in {\mathcal{F}}({{\bsQ}});\ f(\bsQ \bsy) = f(\bsQ \bsy_t)\right\}
\end{equation*} 
with  $f(\bsv) = \argmin_i [\bsv]_i$. Thus we propose a simple proximity test to determine whether \(\bsy_t\) should be retained. More precisely,  we propose to update the set of relevant observations $\bsV_{t-1}$ using  \eqref{eq:updating_rule2_strict} only if
$$
\forall \bsy \in \bsV_{t-1} \cap {\mathcal{A}}({\hat{\bsQ}_{t-1}},\bsy_t),\ \norm{\hat{\bsQ}_{t-1}(\bsy_t-\bsy)}_2 \geq d
$$
i.e., the observation $\bsy_t$ is outside any ball of radius $d$ centered on an observation located on the same facet.

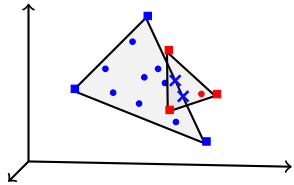
\begin{figure}[b]
    \centering
    \vspace{-0.75cm}
    \begin{tikzpicture}[scale=0.7]
        
            \draw[thick,->] (-1,-0.5,0) -- (4,-0.6,0);  
            \draw[thick,->] (-1,-0.5,0) -- (-1,2.5,0);    
            \draw[thick,->] (-1,-0.5,0) -- (-1,-0.5,1);  

            \coordinate (S1) at (1,2,3);  
            \coordinate (S2) at (3.5,1,3);  
            \coordinate (S3) at (2,3,2);  
            
            \fill[gray,opacity=0.1] (S1) -- (S2) -- (S3) -- cycle; 
            \draw[thick] (S1) -- (S2) -- (S3) -- cycle;

            \filldraw[blue] (S1) ++(-1pt,-1pt) rectangle +(4pt,4pt) node[above left, xshift=10pt, yshift=-10pt] {};
            \filldraw[blue] (S2) ++(-1pt,-1pt) rectangle +(4pt,4pt) node[below right, yshift=10pt] {};
            \filldraw[blue] (S3) ++(-1pt,-1pt) rectangle +(4pt,4pt) node[above, xshift=-4pt] {};

     \foreach \p in {(2.2,2.1,2.6), (1.9,2.7,2.4), (2.5,2.3,2.7), (1.5,2.3,2.7), (0.8,1,0.5),
                     (2.4,1.8,2.1), (1.1,0.6,0), (1.8,0.25,0)  } {
        \filldraw[blue, opacity=1] \p circle (1.5pt);
    }

     \foreach \p in {(2.9,1.7,2.5), (2.75,2.,2.5)} {
        \filldraw[blue, opacity=1] \p node[line width=1pt, cross=3pt,blue] {};
    }


\foreach \p in {  (3.4,1.9,2.9)} {
        \filldraw[red, opacity=1] \p circle (1.5pt);
    }


            \coordinate (S1new) at (2.8,1.6,3);  
            \coordinate (S2new) at (3.7,1.9,3);  
            \coordinate (S3new) at (2.4,2.35,2);  
            
            \fill[gray,opacity=0.1] (S1new) -- (S2new) -- (S3new) -- cycle; 
            \draw[thick] (S1new) -- (S2new) -- (S3new) -- cycle;

            \filldraw[red] (S1new) ++(-1pt,-1pt) rectangle +(4pt,4pt) node[above left, xshift=10pt, yshift=-17pt] {};
            \filldraw[red] (S2new) ++(-1pt,-1pt) rectangle +(4pt,4pt) node[below right, yshift=10pt, xshift=2pt] {};
            \filldraw[red] (S3new) ++(-1pt,-1pt) rectangle +(4pt,4pt) node[above, xshift=10pt, yshift=-6pt] {};

        \end{tikzpicture}
    \caption{Degeneracy of the problem illustrated with $K=3$. Blue dots are relevant observations at iteration $t-1$, while the red dot is the new  observation $\bsy_t$. By imposing strict updating rules, the estimated simplex at iteration $t$ (whose vertices are the red squares) only enclose two relevant observations located on one facet (blue crosses) and the new observation $\bsy_t$.}
    \label{fig:illconditioning}
\end{figure}

\subsection{Practical implementation}
\label{sec: pseudocode}
From a practical implementation point-of-view, the strict updating rules \eqref{eq:updating_rule1_strict} and \eqref{eq:updating_rule2_strict} proposed to update the set of relevant observations $\bsV_{t-1}$ may lead to high sensibility of the MVCU algorithm with respect to (w.r.t.) possible noise degrading the observations. In even more particular scenarios for instance illustrated in Fig.~\ref{fig:illconditioning}, when only a few observations lie on the facets, namely the blue crosses, the MVCU problem under consideration may depart significantly from the target MVCU problem $\mathcal{P}_T$. This may result in searching for a simplex enclosing only a few observations in \(\bsY_{\leq t}\). To address these challenging scenarios, we propose to relax these strict updating rules by selecting relevant observations as those that are close to violating the constraints or close to the facets. This can be achieved by replacing the sets defining the simplex and its facets in \eqref{eq:updating_rule1_strict} and \eqref{eq:updating_rule2_strict} by some \emph{robust} counterparts 
$$\mathcal{S}_{\boldsymbol{\epsilon}}(\bsQ)  = \left\{\bsy \in \mathbb{R}^K;\ g(\bsQ \bsy_j) + \epsilon_1 \ge 0 \; \text{and} \;  |h(\bsQ \bsy_j)| \leq \epsilon_2 \right\}$$
$$\text{and} \qquad \mathcal{F}_{\boldsymbol{\epsilon},\eta}(\bsQ) = \left\{\bsy \in \mathcal{S}_{\boldsymbol{\epsilon}}(\bsQ);\ g(\bsQ \bsy_j) \leq \eta \right\}$$ 
defined by positive tolerance parameters $\boldsymbol{\epsilon}=[\epsilon_1,\epsilon_2]$ and $\eta$. Besides, it is worth noting that the so-called \emph{robust} simplex $\mathcal{S}_{\boldsymbol{\epsilon}}(\bsQ_{t-1})$ is the one actually used to check if the new observation $\bsy_t$ violates the constraint and thus calls for updating $\hat{\bsQ}_{t-1}$ as detailed in Section \ref{sec: naive}.

An algorithmic sketch of one iteration $t$ of the proposed oSSMF method is provided in Algo.~\ref{alg:onlineSISAL}. When a new observation \(\bsy_t\) is available, the first step consists in checking either it belongs to the current estimated (robust) simplex $\mathcal{S}_{\boldsymbol{\epsilon}}(\hat{\bsQ}_{t-1})$ (see line \ref{line:check1}). If it violates one of the underlying constraints (i.e., $g(\hat{\bsQ}_{t-1}\bsy_t)+\epsilon_1<0$ or $|h(\hat{\bsQ}_{t-1}\bsy_t)|>\epsilon_2$), the simplex vertices are updated by running the off-the-shelf MVCU algorithm (see line \ref{line:runMVCU}) and the set of relevant observations is updated (see line \ref{line:update_relevant1}). Otherwise, no update of the simplex is required, and the observation \(\bsy_t\) is considered as a relevant observation only if it is close to a facet but sufficiently distant from other relevant observations (see line \ref{line:update_relevant2a}).

\begin{algorithm}[t]
\caption{\textsf{oSSMF }}\label{alg:onlineSISAL}
\SetKwInOut{KwIn}{Input}
\SetKwInOut{KwOut}{Output}
    \KwIn {new observation $\bsy_t$, current estimate $\Hat{\bsQ}_{t-1}$, current set of relevant observations $\bsV_{t-1}$, proximity parameters $\epsilon_1$, $\epsilon_2$, $d$ and $\eta$.}
        \mycomment{Check either the observation $\bsy_t$ violates the constraints}
        \eIf{ $\bsy_t \notin \mathcal{S}_{\boldsymbol{\epsilon}}(\hat{\bsQ}_{t-1})$ \label{line:check1}}{ 
        \mycomment{Run the \textnormal{MVCU} algorithm}
        \( \Hat{\bsQ}_t  \gets \textsf{MVCU}(\bsV_{t-1} \cup \{\bsy_t\} ,\Hat{\bsQ}_{t-1})\) \label{line:runMVCU} \\       
        \mycomment{Update the set of relevant observations}
        \(\bsV_t \gets \left\{\bsV_{t-1} \cup \left\{\bsy_t\right\}\right\} \cap \left\{\mathcal{S}_{\boldsymbol{\epsilon}}(\hat{\bsQ}_t)^\mathrm{c} \cup \mathcal{F}_{\boldsymbol{\epsilon}, \eta}(\hat{\bsQ}_t) \right\}\)   \label{line:update_relevant1}   
        }
        {
        \mycomment{The observation $\bsy_t$ satisfies the constraints}
        \( \Hat{\bsQ}_t \gets \Hat{\bsQ}_{t-1} \)\\  
            \mycomment{Check either the observation $\bsy_t$ is on/near a facet}
            \mycomment{and update the set of relevant observations accordingly}
            \If{ $\bsy_t \in \mathcal{F}_{\boldsymbol{\epsilon}, \eta}(\hat{\bsQ}_t)$}
            {
            \mycomment{Extract the observations on/near the same facet}
            \(\bsF \gets \bsV_{t-1} \cap {\mathcal{A}}_{\boldsymbol{\epsilon}, \eta}({\hat{\bsQ}_{t}},\bsy_t) \)\\
            \mycomment{Check if $\bsy_t$ is far from these observations}
            \If{$\min_{\bsy \in \bsF} \|\Hat{\bsQ}_{t}(\bsy_t - \bsy)\|_2 \geq d$}
             {\(\bsV_t \gets \bsV_{t-1} \cup \{ \bsy_t\} \) \label{line:update_relevant2a} \\
             }              
            } 
        }
    \KwOut{\( \Hat{\bsQ}_t \) and $\bsV_t$. }
    \end{algorithm}

\section{Experiments}

\noindent \emph{Baseline MVCU and compared methods.} \, The proposed oSSMF is designed to embed any MVCU algorithm solving SSMF (see line \ref{line:runMVCU} of Algo. \ref{alg:onlineSISAL}). Among the available MVCU algorithms, simplex identification via split augmented Lagrangian (SISAL) has received widespread attention thanks to its remarkable estimation accuracy, its robustness to noise and outliers, as well as its relatively low computational burden \cite{bioucas2009variable,huang2022sisal}. Hence, we implement the proposed oSSMF with SISAL, which is also used as an MVCU baseline. The goal is to assess how closely the performance of oSSMF matches that of the canonical implementation of SISAL. To place SISAL under optimal conditions, it is adapted to the online context by running it at each iteration \( t \) on the whole set \(\bsY_{\leq t}\) of  available observations so far. We also compare oSSMF to the state-of-the-art online dictionary learning (ODL) method \cite{mairal2010online} as well as the recent KF--OSU algorithm \cite{kouakou2024fly}.\\ \vspace{-0.20cm}

\noindent \emph{Implementation details.} \, All methods have been implemented in MATLAB R2022b, on a laptop equipped with an Intel Core i7-8565U processor (1.80 GHz) and 8 GB of RAM. The proposed oSSMF method has been executed with the parameters \(\epsilon_1 = \epsilon_2 = 0.0001\), \(\eta = 0.03\) and \( d = 0.7\), with an online subspace estimation algorithm \cite{arora2012stochastic} which updates the signal subspace iteratively. The choice and impact of the algorithmic parameters are further discussed in Appendix \ref{app:parameters}.\\ \vspace{-0.20cm}

\noindent \emph{Synthetic (\textsf{SD}) and real (\textsf{RD}) datasets.} \, 200 synthetic datasets were generated according to the model (\ref{model}) with \(L = 400\), \(T = 10000\), and \(K = 7\). 
The basis vectors in \(\bsS\) have been constructed by summing Gaussian functions with varying means, amplitudes, and standard deviations. The columns of \(\bsC\) have been drawn from a truncated uniform distribution over the simplex, with a maximum purity level of $0.7$  \cite{abdolali2021simplex}. White Gaussian noise has been added to reach a signal-to-noise ratio (SNR) of $15$dB.
Besides, a \(101 \times 101 \times 343\) real Raman hyperspectral image (HI) was acquired over a tablet composed of NaNO\(_3\), CaCO\(_3\), and Na\(_2\)SO\(_4\) powders, using a LabRAM HR microspectrometer \cite{coic2023assessment}. The HI yields a \(343 \times 10201\) matrix, from which 199 row-shuffled replicas are generated to test robustness to observation order.\\ \vspace{-0.20cm}

\noindent \emph{Performance metrics.} \, Two standard metrics have been considered to assess the performance of the compared methods~\cite{miao2007endmember, bioucas2010alternating}: the average spectral angle distance (in degrees)  defined as \(
\text{aSAD} = \frac{1}{K} \sum_{k=1}^{K} \arccos\left( \frac{\bss_k^\top \hat{\bss}_k}{\norm{\hat{\bss}}_2\norm{\bss}_2} \right),
\) and the root-mean-square error  defined as \(
\text{RMSE} = \sqrt{\frac{1}{KN} \| \bsC - \Hat{\bsC} \|^2_\text{F}},\) where the mixing coefficients $\bsC$ at iteration $t$ have been estimated by the variable splitting and augmented Lagrangian (SUnSAL) algorithm \cite{bioucas2010alternating}, using the estimated basis vectors $\hat{\bsS}_t$ and the full set of observations \(\bsY\). The methods are also compared w.r.t. the processing time per iteration.\\ \vspace{-0.20cm}

\begin{figure}[h!]
  \centering
\input{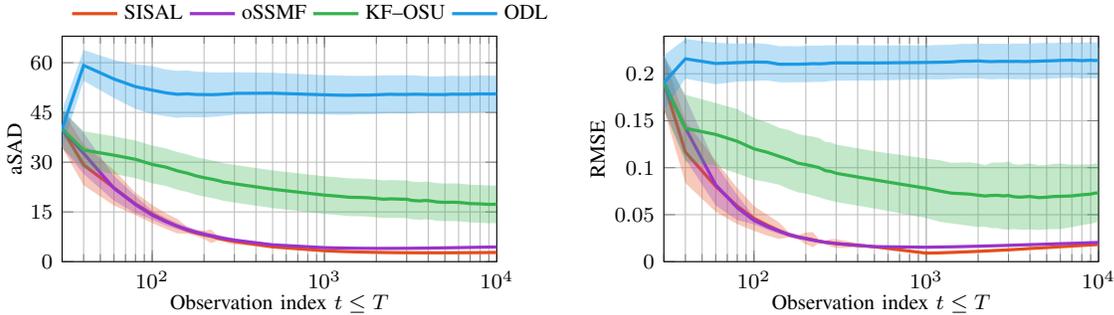}
\caption{Performance of the methods on \textsf{SD}, averaged over the 200 datasets.}
  \label{fig:syntheticPerf}
\end{figure}

\noindent \emph{Results and discussion.} \,
Fig.~\ref{fig:syntheticPerf} shows the evolution of aSAD and RMSE over iterations on the dataset \textsf{SD}. Overall, SISAL yields the lowest values for both metrics, with limited fluctuations, indicating stable estimation. oSSMF follows a similar trend with slight deviations, suggesting comparable performance. KF--OSU performs less favorably, with noticeably higher aSAD and RMSE than oSSMF while ODL reports the highest values for aSAD and RMSE.

\begin{figure}[h!]
  \centering
\input{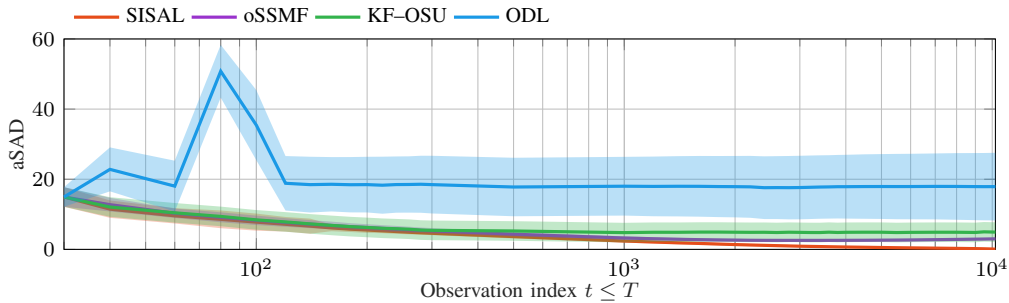}
\caption{Performance of the methods on \textsf{RD}, averaged over $200$ row-shuffled replicas of the original dataset.}
  \label{fig:realPerf}
\end{figure}

Fig.~\ref{fig:realPerf} presents the aSAD over iterations on the data set \textsf{RD}, computed w.r.t. reference vertices extracted by SISAL from the full set $\bsY$ of $T=10201$ observations. SISAL and oSSMF achieve comparable values. KF--OSU and ODL yield lower performance than oSSMF, with ODL exhibiting the largest fluctuations around a significantly higher aSAD value.

Table~\ref{tab:runtime} summarizes the computational times per iteration required by the compared algorithms. SISAL is the slowest. oSSMF provides a speed-up of approximately 28× and 54× on the datasets \textsf{SD} and \textsf{RD}, respectively, with reduced variability. ODL yields the lowest average runtime, followed by KF--OSU. These experimental results are further discussed in Appendix \ref{app:numerical}.

\begin{table}[h!]
\centering
\caption{Average runtime per iteration (in seconds) for each method.}
\renewcommand{\arraystretch}{1.1}
\setlength{\tabcolsep}{4pt}
\begin{tabular}{ccccc}
\toprule
       & SISAL           & oSSMF & KF--OSU & ODL\footnotemark \\
\hline
\textsf{SD}     & $0.25$ \scriptsize{$\pm 0.113$} & $0.009$  \scriptsize{$ \pm 0.008$} & $0.009 $  \scriptsize{$\pm 0.0001$} & $0.002 $  \scriptsize{$\pm 0.0001$} \\
\textsf{RD}     & $0.163$  \scriptsize{$\pm 0.09$} & $0.003 $  \scriptsize{$\pm 0.002$} & $0.005 $  \scriptsize{$\pm 0.0001$} & $0.0004 $  \scriptsize{$\pm 0.0001$} \\
\bottomrule
\end{tabular}
\label{tab:runtime}
\end{table}
\footnotetext{The number of iterations, a tunable parameter in ODL, was set to 1, as this value yielded the best performance in our experiments.}

\section{Conclusion}

This work introduces oSSMF, an online extension of the MVCU algorithm tailored for online SSMF. By leveraging the problem constraints, oSSMF efficiently updates the basis vectors incrementally without requiring full data access, using an online selection strategy to identify informative observations. Experimental results demonstrate substantial speed gains over the canonical implementation of the MVCU algorithm SISAL without sacrificing the estimation performance.

\appendices

\section{Impact and selection of algorithmic parameters}\label{app:parameters}

As exhibited in Algo.~$1$, the selection of the relevant observations on which the computational burden of oSSMF depends is parametrized by the algorithmic parameters $\epsilon_1$, $\epsilon_2$, $d$ and $\eta$. Specifically, the parameters $\epsilon_1$ and $\epsilon_2$ regulate how strictly the problem constraints are enforced when selecting essential observations and thus governs the decision to update the currently estimated simplex (i.e., to run the MVCU algorithm or not). With $\epsilon_1 = \epsilon_2 = 0$, updates occur whenever a new observation lies outside the simplex. For $(\epsilon_1, \epsilon_2) \neq (0,0)$, a tolerance margin is introduced, reducing unnecessary updates when the current estimated simplex (vertices) already provides a good approximation or to mitigate the effect of measurement noise. The choice $\epsilon_1 = \epsilon_2 = 10^{-4}$ was found effective when conducting the experiments reported in this paper.  

In its turn, the parameter $d \in [0,1]$ controls redundancy among relevant observations: as $d \to 0$, redundant observations may tend to be included in the estimation procedure; as $d \to 1$, redundancy between close observations is reduced extensively, with the risk of discarding important observations. Empirically, $d \in [0.55,\,0.75]$ yielded a reasonable trade-off between estimation accuracy and computational efficiency.  

Finally, the parameter $\eta \in [0,1]$ serves to prevent (or at least limit) degeneracy of the problem. As schematically illustrated in Fig.~\ref{fig:illconditioning}, a too small threshold may lead to selecting too few observations near the facets, resulting in poorer and unreliable estimates (i.e., estimates significantly deviating from the current simplex). Conversely, setting $\eta$ close to $1$ avoids this issue, at the price of incorporating more observations in $\bsV_t$ than necessary, resulting in a significant increase of the computational burden. We found empirically that $\eta=0.03$ was a reasonable choice. Note that this choice may depend on the geometric structure of the data set at each iteration $t$, which is a priori unknown. A possible direction for future work is to adaptively adjust $\eta$ on the fly as new observations are acquired.

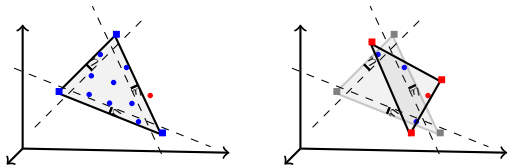
\begin{figure}[h]
\centering
{%
\begin{tikzpicture}[scale=0.55, every node/.style={inner sep=0,outer sep=0}]
  \draw[thick,->] (-1,-0.5,0) -- (4,-0.6,0);  
  \draw[thick,->] (-1,-0.5,0) -- (-1,2.5,0);  
  \draw[thick,->] (-1,-0.5,0) -- (-1,-0.5,1); 

  \coordinate (S1) at (1,2,3);  
  \coordinate (S2) at (3.5,1,3);  
  \coordinate (S3) at (2,3,2);  
            
  \fill[gray,opacity=0.1] (S1) -- (S2) -- (S3) -- cycle; 
  \draw[thick] (S1) -- (S2) -- (S3) -- cycle;

  \def\xoffsetOne{0.45} 
  \coordinate (S1Line) at (1 +\xoffsetOne,2,3);
  \coordinate (S2Line) at (3.5 +\xoffsetOne,1,3);
  \draw[dashed] ($(S1Line)!-0.6!(S2Line)$) -- ($(S1Line)!1.3!(S2Line)$); 
  \coordinate (M)  at ($(S1)!0.5!(S2)$);  
  \coordinate (F) at ($(S1Line)!(M)!(S2Line)$);
  \draw[thick] (M) -- (F); 
  \draw[thick] (M) -- node[pos=.4, sloped, below=1pt] {\tiny$\eta$} (F);

  \def\xoffsetTwo{0.28}
  \coordinate (S1Line) at (1 +\xoffsetTwo,2,3);
  \coordinate (S3Line) at (2 +\xoffsetTwo,3,2);         
  \draw[dashed] ($(S1Line)!-0.6!(S3Line)$) --  ($(S1Line)!1.3!(S3Line)$); 
  \coordinate (M)  at ($(S1)!0.5!(S3)$);  
  \coordinate (F) at ($(S1Line)!(M)!(S3Line)$);
  \draw[thick] (M) -- (F); 
  \draw[thick] (M) -- node[pos=.4, sloped, above=1pt] {\tiny $\eta$} (F);

  \def\xoffsetThree{0.21}
  \coordinate (S2Line) at (3.5 -\xoffsetThree,1,3);
  \coordinate (S3Line) at (2 -\xoffsetThree,3,2);         
  \draw[dashed] ($(S2Line)!-0.2!(S3Line)$) --  ($(S2Line)!1.3!(S3Line)$);       
  \coordinate (M)  at ($(S2)!0.5!(S3)$);  
  \coordinate (F) at ($(S2Line)!(M)!(S3Line)$);
  \draw[thick] (M) -- (F); 
  \draw[thick] (M) -- node[pos=.4, sloped, below=1pt] {\tiny $\eta$} (F);
          
  \filldraw[blue] (S1) ++(-1pt,-1pt) rectangle +(4pt,4pt);
  \filldraw[blue] (S2) ++(-1pt,-1pt) rectangle +(4pt,4pt);
  \filldraw[blue] (S3) ++(-1pt,-1pt) rectangle +(4pt,4pt);

  \foreach \p in {(2.2,2.1,2.6), (1.8,2.7,2.4), (2.55,2.5,2.7), (1.7,2.3,2.7), (0.8,1,0.5),
                  (2.45,1.4,2.1), (1.1,0.6,0), (1.8,0.15,0)} {
    \filldraw[blue, opacity=1] \p circle (1.5pt);
  }

  \foreach \p in {(3.2,1.9,2.9)} {
    \filldraw[red, opacity=1] \p circle (1.5pt);
  }
\end{tikzpicture}}\qquad %
{%
\begin{tikzpicture}[scale=0.55, every node/.style={inner sep=0,outer sep=0}]
  \draw[thick,->] (-1,-0.5,0) -- (4,-0.6,0);  
  \draw[thick,->] (-1,-0.5,0) -- (-1,2.5,0);  
  \draw[thick,->] (-1,-0.5,0) -- (-1,-0.5,1); 

  \coordinate (S1) at (1,2,3);  
  \coordinate (S2) at (3.5,1,3);  
  \coordinate (S3) at (2,3,2);  
            
  \fill[gray,opacity=0.1] (S1) -- (S2) -- (S3) -- cycle; 
  \draw[thick, lightgray] (S1) -- (S2) -- (S3) -- cycle;

  \def\xoffsetOne{0.45} 
  \coordinate (S1Line) at (1 +\xoffsetOne,2,3);
  \coordinate (S2Line) at (3.5 +\xoffsetOne,1,3);         
  \draw[dashed] ($(S1Line)!-0.6!(S2Line)$) --  ($(S1Line)!1.3!(S2Line)$); 
  \coordinate (M)  at ($(S1)!0.5!(S2)$);  
  \coordinate (F) at ($(S1Line)!(M)!(S2Line)$);
  \draw[thick] (M) -- (F); 
  \draw[thick] (M) -- node[pos=.4, sloped, below=1pt] {\tiny $\eta$} (F);

  \def\xoffsetTwo{0.28}
  \coordinate (S1Line) at (1 +\xoffsetTwo,2,3);
  \coordinate (S3Line) at (2 +\xoffsetTwo,3,2);         
  \draw[dashed] ($(S1Line)!-0.6!(S3Line)$) --  ($(S1Line)!1.3!(S3Line)$); 
  \coordinate (M)  at ($(S1)!0.5!(S3)$);  
  \coordinate (F) at ($(S1Line)!(M)!(S3Line)$);
  \draw[thick] (M) -- (F); 
  \draw[thick] (M) -- node[pos=.4, sloped, above=1pt] {\tiny $\eta$} (F);

  \def\xoffsetThree{0.21}
  \coordinate (S2Line) at (3.5 -\xoffsetThree,1,3);
  \coordinate (S3Line) at (2 -\xoffsetThree,3,2);         
  \draw[dashed] ($(S2Line)!-0.2!(S3Line)$) --  ($(S2Line)!1.3!(S3Line)$);       
  \coordinate (M)  at ($(S2)!0.5!(S3)$);  
  \coordinate (F) at ($(S2Line)!(M)!(S3Line)$);
  \draw[thick] (M) -- (F); 
  \draw[thick] (M) -- node[pos=.4, sloped, below=1pt] {\tiny $\eta$} (F);
          
  \filldraw[gray] (S1) ++(-1pt,-1pt) rectangle +(4pt,4pt);
  \filldraw[gray] (S2) ++(-1pt,-1pt) rectangle +(4pt,4pt);
  \filldraw[gray] (S3) ++(-1pt,-1pt) rectangle +(4pt,4pt);

  \foreach \p in {(1.8,2.7,2.4), (2.55,2.5,2.7), (1.8,0.15,0)} {
    \filldraw[blue, opacity=1] \p circle (1.5pt);
  }

  \foreach \p in {(3.2,1.9,2.9)} {
    \filldraw[red, opacity=1] \p circle (1.5pt);
  }

  \coordinate (S1new) at (2.8,1.,3);  
  \coordinate (S2new) at (1.85,3.2,3);  
  \coordinate (S3new) at (3.15,1.9,2);  
  \fill[gray,opacity=0.1] (S1new) -- (S2new) -- (S3new) -- cycle; 
  \draw[thick] (S1new) -- (S2new) -- (S3new) -- cycle;
  \filldraw[red] (S1new) ++(-1pt,-1pt) rectangle +(4pt,4pt);
  \filldraw[red] (S2new) ++(-1pt,-1pt) rectangle +(4pt,4pt);
  \filldraw[red] (S3new) ++(-1pt,-1pt) rectangle +(4pt,4pt);
\end{tikzpicture}}%
\caption{Influence of the parameter $\eta$ on the simplex vertices (squares) estimated at iteration $t$ (left) and at iteration $t+1$ (right). Relevant observations at iteration $t$ are shown as blue dots, while a new observation that violates the constraint is shown as a red dot. At iteration $t+1$, the simplex vertices (red squared) recovered from relevant observations  near the facets (blue dots) and the new observation (red dot) signficantly departs from the previous estimates.}
\label{fig:illconditioning}
\end{figure}

\section{Detailed analysis of the numerical results}\label{app:numerical}

Fig.~$2$ in the main part of the paper reports the performance of the compared methods obtained on the synthetic datasets (\textsf{SD}) in terms of aSAD and RMSE as a function of the observation index. SISAL exhibits good performance, as indicated by low aSAD and RMSE values, particularly from time index $t=300$ onward. Small deviations around the mean can be observed for both metrics, suggesting reliable estimates. oSSMF shows similar trends to SISAL, with minor deviations, demonstrating comparable estimation accuracy and reliability. The small gap of performance between SISAL and oSSMF may be attributed to potential errors in the signal subspace estimation and/or the inappropriate discarding of some relevant observations for some iterations.  KF--OSU performs less effectively, with a noticeable gap from oSSMF in both aSAD and RMSE. Finally, ODL performs poorly, with high values of aSAD and RMSE, despite its fast convergence. This can be attributed to the fact that ODL does not exploit the geometric constraints specific to SSMF (namely non-negativity and sum-to-one).

Fig.~$3$ illustrates the performance of the compared methods on the real dataset (\textsf{RD}) in term of aSAD. In absence of associated ground truth for this real data set, aSAD is computed to measure the spectral discrepancy from a reference estimated by SISAL when it is applied on the full set of $T=10201$ observations. The results show nearly similar performance between SISAL and oSSMF. In contrast, KF--OSU and ODL exhibit overall lower performance and higher fluctuations around the mean value. This behavior suggests a less reliable estimation by KF--OSU and ODL. Finally, although ODL underperforms oSSMF, the gap is smaller on \textsf{RD} compared to \textsf{SD}, which may be explained by the lower rank of the problem ($K=3$ for \textsf{RD} vs. $K=7$ for \textsf{SD}) as well as the high SNR of RD. 

\bibliographystyle{IEEEtran}
\bibliography{strings_all_ref,biblio}

\end{document}